# HIT RATIO DRIVEN MOBILE EDGE CACHING SCHEME FOR VIDEO ON DEMAND SERVICES

*Xing Chen, Lijun He, Shang Xu, Shibo Hu, Qingzhou Li, Guizhong Liu*[*]

The Department of Electronic and Information Engineering, Xi'an Jiao Tong University, Xi'an, China

## ABSTRACT

More and more scholars focus on mobile edge computing (MEC) technology, because the strong storage and computing capabilities of MEC servers can reduce the long transmission delay, bandwidth waste, energy consumption, and privacy leaks in the data transmission process. In this paper, we study the cache placement problem to determine how to cache videos and which videos to be cached in a mobile edge computing system. First, we derive the video request probability by taking into account video popularity, user preference and the characteristic of video representations. Second, based on the acquired request probability, we formulate a cache placement problem with the objective to maximize the cache hit ratio subject to the storage capacity constraints. Finally, in order to solve the formulated problem, we transform it into a grouping knapsack problem and develop a dynamic programming algorithm to obtain the optimal caching strategy. Simulation results show that the proposed algorithm can greatly improve the cache hit ratio.

## 1. INTRODUCTION

With the full development of the fifth generation mobile communication research, mobile communication technology and industry is entering a new stage of development. It will meet the needs of applications with ultra-high traffic density, ultra-high connection density and ultra-high mobility. Therefore, ultra-high-definition video, virtual reality, augmented reality, online games and other extreme services are going to be provided for users. As a main evolutional technology in 5G communication system, MEC [1][2] is close to user nodes and data sources. The problems of long transmission delay, Bandwidth waste, energy consumption and privacy leak are avoided by its powerful intelligent storage and computing capabilities [3][4]. Therefore, by pre-storing some popular videos in the cache of MEC, users can retrieve their request videos mainly from the MEC cache. As for other uncached videos, users can retrieve them from the Internet video server via the constrained backhaul for video on demand services. MEC caching not only reduces users' startup delay, but also reduces the backhaul network load. In this context, the main task of our work is to simultaneously reduce the startup delay and backhaul load by designing a caching placement strategy.

The idea of caching has been around for a long time. The popular videos were previously cached on a proxy or small base station with storage capabilities. Literature [5] proposed a novel cache architecture with a proxy server. This architecture allows partial caching of media objects and joint delivery from caches and origin servers to reduce startup delay and improve stream quality. In 5G wireless networks, caching some popular content at femto base-stations (FBSs) and user equipment (UE) can be exploited to alleviate the burden of backhaul and to reduce the costly transmissions from the macro base-stations to UEs [6][7].

With the advent of MEC technology, MEC caching is used in many applications with strict delay requirements. A joint optimization scheme was proposed in wireless cellular networks with mobile edge computing, taking into consideration computation offloading decision, physical spectrum resource allocation, MEC computation resource allocation, and content caching strategy [8]. Literature [9] proposed a storage resource allocation scheme of the MEC server taking each BS traffic load into consideration. The caching scheme not only considers the popularity of the videos but also wireless channel conditions. Literature [10] studied a QoE driven mobile edge caching placement optimization problem for dynamic adaptive video streaming that properly takes into account the different rate-distortion(R–D) characteristics of videos and the coordination among distributed edge servers. The proposed algorithm guides the edge server to select the cached representations in practice based on both the video popularity and content information. Literature [11] jointly considered Mobile Edge Computing and Caching-enabled software-defined mobile networks to enhance the video service in next generation mobile networks. Literature [12] presented a novel caching replacement algorithm named Weighted Greedy Dual Size Frequency (WGDSF) algorithm, which is an improvement on the Greedy Dual Size Frequency (GDSF) algorithm. The WGDSF algorithm

---

mainly adds weighted frequency-based time and weighted document type to GDSF for improving the caching hit ratio.

The above mentioned literatures adopt several caching algorithms and optimization objectives. The most popular caching algorithms are Least Recently Used (LRU) and the Least Frequently Used (LFU) because of their implementation simplicity. The most popular optimization objectives are to reduce the backhaul network load and startup delay. In this paper, we construct a caching scheme that accounts for video popularity, user's content preference and video representation to improve the cache hit rate of all the users under the MEC server. The contributions of this paper can be summarized as follows:

1) More comprehensive request probability calculation.

    Existing cache algorithms only consider video popularity and representation characteristics, without considering the video content preference of each user, which results in a low cache hit rate. However, as a very important feature in the video on demand services, user preference can better characterize the request probability of videos. Therefore, we jointly consider the video popularity, the user preference, and the characteristic of video representations into the request probability calculation to guide a differential caching.

2) Hit ratio driven joint optimization of caching and real-time transcoding.

    In practice, the main purpose of MEC server is to enable more users to enjoy better video on demand services. Existing works mainly aimed at minimizing one of the factors: backhaul network load and or startup delay, which cannot achieve an optimal realistic purpose. To deal with the issue, we simultaneously improve the above objectives with the objective to maximize the cache hit ratio, and combine real-time transcoding to enable more users to enjoy better services. Correspondingly, we propose a hit ratio driven joint optimization problem of caching and real-time transcoding. To solve this problem, we transform it into a grouping knapsack problem and the dynamic programming algorithm is applied to obtain the optimal solution.

The rest of the paper is organized as follows: Section II shows the network topology and the system framework, moreover the characteristics of each network module is illustrated. Section III presents the caching problem and the mathematical model. The problem solution is given in section IV. Section V provides the simulation results. Finally, Section VI concludes this paper.

## 2. SYSTEM MODEL

In this paper, we consider a scenario where $N$ users can request different representations of different videos from the cloud server or MEC server for video on demand application in MEC system, as shown in Figure 1. The MEC system is mainly composed of cloud server, MEC server, users and base stations. The cloud server is the data center to provide users with all the videos. The MEC server has powerful storage and computing capacity, which are used to cache and transcode videos, respectively. Since the required processing capacity for transcoding is small compared to handling offloading tasks, the computing capacity for transcoding is assumed to be unlimited in this work. If the representation of the video is cached in MEC server, it will be directly obtained from the MEC server. If the higher representation of the video is cached in the MEC server, the lower representation of the video is obtained through the transcoding function of MEC server. Otherwise it will be obtained from the cloud server through backhaul network.

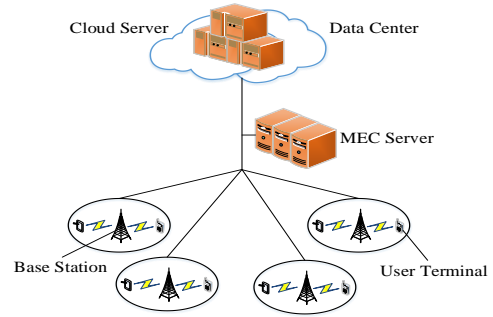

Fig.1 System model

We denote the MEC server, the set of base stations and the set of users by $\Gamma$, $b=\{1,2,\cdots,M\}$, $u=\{1,2,\cdots,N\}$, respectively. The request arrival time of these users is a Poisson distribution with intensity $\lambda$. There are $K$ videos, denoted by $v=\{1,2,\cdots,K\}$, and each video is encoded into two representations ($l \in \{1,2\}$): the High-Definition video ($l=1$) and the Standard-Definition video ($l=2$). The size of representation $l$ of video $v$ is denoted by $s_v^l$. These videos are divided into several types by content (i.e. news, sport, person, etc.). Based on video popularity, user preference, and representation characteristic, we can estimate the probability that each user requests each representation of each video.

(1) *Video popularity*: We arrange videos in descending order of popularity with the number $f \in \{1,2,\cdots,K\}$; $f_v=1$ indicates that the v-th video has the highest popularity, $f_v = K$ indicates that the v-th video has the lowest popularity, and the video popularity follows a Zipf distribution with the decay parameter $\gamma$. We consider that the video popularity varies periodically. The request probability of the video $v$ is calculated by

$$r_v = f_v^{-\gamma} \bigg/ \sum_{v=1}^{K} f_v^{-\gamma} \qquad (1)$$

(2) *User preference*: We consider that each user has one preference video type. The preference level is defined by the preference factor $\alpha_u \in \{0.2, 0.5, 0.8\}$. The greater the value of $\alpha_u$, the higher the level of the preference. If the content of video $v$ is the same as the user's preference type, $r_{v,u}$ is obtained by

$$r_{v,u} = r_v \cdot (1 + \alpha_u) \qquad (2)$$

Conversely, if the content of video $v$ is different from the user's preference type, $r_{v,u}$ is obtained by

$$r_{v,u} = r_v \cdot (1 - \alpha_u) \qquad (3)$$

Then, the request probability $\hat{r}_{v,u}$ is obtained by normalization $r_{v,u}$, as follows:

$$\hat{r}_{v,u} = r_{v,u} \bigg/ \sum_{v=1}^{K} r_{v,u} \qquad (4)$$

(3) *The characteristic of video representations*: Let $g(f_v)$ and $1 - g(f_v)$ denote the characteristic for the Standard-Definition(SD) and the High-Definition(HD) qualities for the video $v$, respectively [14], where $g(f_v) \in [0,1]$ is calculated by

$$g(f_v) = (f_v - 1)/(K - 1) \qquad (5)$$

The main idea is that the higher popularity of the video, the more users who request this video. Therefore, the high popularity HD video is caching, that can improve the viewing experience of most users.

The characteristic of representations is considered into $\hat{r}_{v,u}$, and the request probability of HD video is calculated by

$$\hat{r}_{v,u}^1 = \hat{r}_{v,u} \cdot (1 - g(f_v)) \qquad (6)$$

The request probability of SD video is calculated by

$$\hat{r}_{v,u}^2 = \hat{r}_{v,u} \cdot g(f_v) \qquad (7)$$

The caching strategy is guided by analyzing the list of users' request probability. From the perspective of the MEC server, the total request probability is obtained for each representation of each video, $req_v^l$. The $req_v^l$ is calculated by

$$req_v^l = \sum_{u=1}^{N} \hat{r}_{v,u}^l \bigg/ N \qquad (8)$$

## 3. JOINT OPTIMIZATION PROBLEM OF VIDEO CACHING AND TRANSCODING

A higher cache hit rate means that more users get videos from the MEC server, so the startup delay of those users and the backhaul network load can be reduced, and a better viewing experience for these users is obtained. In this section, we seek to design the caching strategy that considering real-time transcoding to maximize the caching hit ratio. The caching result is defined by $X = \{x_v^l \in \{0,1\} \mid v = 1, 2, \cdots K; l = 1, 2, \cdots L\}$. Case $x_v^l = 1$ denote that representation $l$ of video $v$ is cached on the MEC server and $x_v^l = 0$ otherwise. The transcoding result is defined by $Y = \{y_v^l \in \{0,1\} \mid v = 1, 2, \cdots K; l = 1, 2, \cdots L\}$. The case $y_v^l = 1$ denote that representation $l$ of video $v$ is obtained by transcoding the higher representation of video $v$ in MEC server and $y_v^l = 0$ otherwise. The total request times is denoted by $Q$ for all the users in a constant period. The total storage of MEC server is denoted by $S$. Therefore, the caching hit ratio is calculated in a constant period by

$$R(X,Y) = \frac{\sum_{v=1}^{K}\sum_{l=1}^{L}\left((x_v^l + y_v^l) \cdot req_v^l \cdot Q\right)}{Q} \qquad (9)$$

If one requested video is obtained from the MEC server or transcoded by it, this request can be regarded as a caching hit. Formally, the caching problem can be formulated as follows:

$$\begin{aligned}
\max \; & R(X,Y) \\
s.t. \quad C1: & \sum_{v=1}^{V}\sum_{l=1}^{L} x_v^l s_v^l \leq S \\
C2: & \sum_{l=1}^{L} x_v^l \leq 1 \qquad (10) \\
C3: & \; y_v^l = \sum_{m=l+1}^{L} x_v^m \\
C4: & \; x_v^l \in \{0,1\}
\end{aligned}$$

The first constraint (C1) ensures that the data size of the cached videos cannot exceed the MEC server caching capacity. Constraint (C2) means that only one representation can be cached for each video. The reason is that the transcoding ability of the MEC server can be used to transcode the video to obtain the requested representation of video, saving more space to cache more videos. Constraint (C3) implies that the lower representation of the video can be obtained by transcoding the higher representation of the video. The reason is that the MEC server has powerful computing ability, whose specific explanation is shown in Section II. Finally, Constraint (C4) represents the possible value of the cache result.

## 4. JOINT CACHING ALGORITHM OF VIDEO CACHING AND TRANSCODING

In this section, we will give the solution to the above optimal problem, which are determined by both the video caching and the transcoding. First, the constraint (C3) is substituted into the objective function, and the optimization problem is further simplified as follows:

$$\max \sum_{v=1}^{K} \sum_{l=1}^{L} \left( \left( x_v^l + \sum_{m=l+1}^{L} x_v^m \right) req_v^l \right)$$

$$s.t. \quad C1: \sum_{v=1}^{V} \sum_{l=1}^{L} x_v^l s_v^l \leq S \quad (11)$$

$$C2: \sum_{l=1}^{L} x_v^l \leq 1$$

$$C3: x_v^l \in \{0,1\}$$

Then, we assume the optimization problem as a grouping knapsack problem. All representations of a video are formed as a group, and each representation of the video is a member of the group. The grouping knapsack problem assumes the form:

$$\max \sum_{v=1}^{K} \sum_{l=1}^{L} x_v^l p_v^l$$

$$s.t. \quad C1: \sum_{v=1}^{V} \sum_{l=1}^{L} x_v^l s_v^l \leq S \quad (12)$$

$$C2: \sum_{l=1}^{L} x_v^l \leq 1$$

$$C3: x_v^l \in \{0,1\}$$

where $p_v^l$ is the increased hit ratio by caching representation $l$ of video $v$, which is equivalent to the value of item in the backpack problem. The quantity $p_v^l$ is calculated by

$$p_v^l = \sum_{m=1}^{l} req_v^m \quad (13)$$

The grouping knapsack problem is a special knapsack problem, its characteristics are described as follows: (i) compared with the general backpack problem, grouping backpack problem divide objects into some groups; (ii) during the loading of the backpack, only one object can be placed in each group at most; (iii) determine which object should be put in the backpack so that the total value of all the objects in the knapsack is the maximum. The grouping knapsack problem is solved by using a dynamic programming algorithm. In the process of solving the problem, a state transition matrix needs to be defined, as $trans(v,w)$, which means the maximum value (the hit ratio) of placing the first $v$ group videos into a MEC server of size $w$. The state transition equation is constructed by

$$trans(v,w) = \max \begin{cases} trans(v-1,w), \\ trans(v-1,w-s_v^l) + p_v^l \mid l=1,2,\cdots,L \end{cases} \quad (14)$$

There are two cases in Equation (14):
1) if the size of each video of the v-th group is greater than $w$, then $trans(v,w) = trans(v-1,w)$;

2) if the size of at least one video is smaller than $w$ in the v-th group, then

$$trans(v,w) = \max \left\{ trans(v-1,w-s_v^l) + p_v^l \mid l=1,2,\cdots L \right\}.$$

Based on the above analysis, the caching result is obtained through the dynamic programming. First, the maximum value is obtained through the iterative method when the size of the backpack is $S$, and then reversely determine which video is cached from the beginning of the last group of videos. Our solution procedures are summarized in Algorithm I, which is executed at the beginning of a constant period.

## 5. PERFORMANCE EVALUATION

*A. Parameter settings*

In this section, we present the numerical results that demonstrate the hit ratio improvement achieved by the joint optimization of the video caching and the real-time transcoding. All the simulations are conducted by MATLAB. The simulation experiment adopts a library of 21 videos with their popularity following the Zipf distribution with parameter $\gamma = 0.6$. Each video has two representations ($L$=2): HD and SD. According to the video content, videos are divided into five types: news, scene, sport, traffic and person. The size of each representation of the videos could be different. And, the size range is 3-65 storage units. The video request arriving follows a Poisson distribution with rate $\lambda$=0.9. The experimental environment deployment only considers the situation of one MEC server. The user number $N$ is 900. And, these users are evenly distributed in the range of 4 base stations. We update the video popularity at 0:00. For all videos, we assume that the request times be 900 every day. The algorithms compared in this section are as follows:

**PROPOSED:** joint optimization of video caching and real-time transcoding in this paper.

**LFU:** Least Frequently Used. The idea of this algorithm is that: the more the data was requested in the previous period, the more the data is requested in the later period.

**LRU:** Least Recently Used. The idea of this algorithm is that: if the data has been requested recently, its probability of being requested in the future is also higher.

**WGDSF:** Weighted Greedy Dual Size Frequency [9]. This algorithm is an improvement on the Greedy Dual Size Frequency (GDSF) algorithm, which adds the factors of weighted frequency-based time and weighted document type to GDSF.

*B. Performance evaluation*

We validate the algorithm performance in term of the caching hit ratio under different simulation configurations. The specific experimental results are shown in the following figures.

**Algorithm I:** Hit Ratio Driven Mobile Edge Caching Algorithm

**Input:** the total request probability, $req_v^l$, $v=1,2,\cdots,K$; $l=1,2,\cdots,L$.

**Output:** the caching result $X = \{x_v^l | v=1,2,\cdots,K; l=1,2,\cdots,L\}$.

1  **Begin:**
2  Initialize the state transition matrix $trans(v,w) = 0$, $v=1,2,\cdots K$, $w=1,2,\cdots S$
3  **for** video $v = 1:K$ **do**
4    **for** representation $l = 1:2$ **do**
5      the value $p_v^l$ is calculated by (10).
6    **end for**
7  **end for**
8  **for** $w = 1:S$ **do**
9    Calculate the maximum value of placing the first group of videos, $trans(1,w)$, according to equation (14).
10 **end for**
11 **for** video $v = 2:K$ **do**
12   **for** $w = 1:S$ **do**
13     Calculate the maximum value of placing the first $v$ group of videos, $trans(v,w)$, according to equation (14).
14   **end for**
15 **end for**
16 **for** video $v = K:-1:2$ **do**
     Reverse the running process of step 13, and then judge which video is cached. $X = \{x_v^l | v=2,3,\cdots K; l=1,2,\cdots,L\}$ is obtained.
17 **end for**
18 Judge which representation is cached in the first group of videos. $x_1^l$ is obtained.
19 **return** $X = \{x_v^l | v=1,2,\cdots K; l=1,2,\cdots,L\}$, $trans(K,S)$.

Figure 2(a) shows the total hit ratio of all the users for different algorithm when the storage size of the MEC server is 350 storage units. As shown in this figure, PROPOSED has the best performance in term of the hit ratio and is about 10% higher than the hit ratio of LFU, LRU and WGDSF algorithms. The reason should be that PROPOSED obtains the optimal solution by dynamic programming. The hit ratio of the WGDSF algorithm is slightly higher than the LRU and LFU algorithms. This is because of the fact that the WGDSF algorithm jointly considers the request frequency and the size of videos.

Figure 2(b) shows the hit ratio of all the users for different algorithms in different storage size of the MEC server. Compared with LFU, LRU and WGDSF, PROPOSED has the best performance in term of cache hit ratio. When the storage size of the MEC server is small, the performance of PROPOSED is less improved due to the limited storage capacity. With the increase of the storage size of the MEC server, the hit ratio will be gradually close to 100%. The reason is that more videos could be cached at the MEC server for all the users. The hit ratios of the three algorithms are 100% when the MEC server could cache high representation of all the videos.

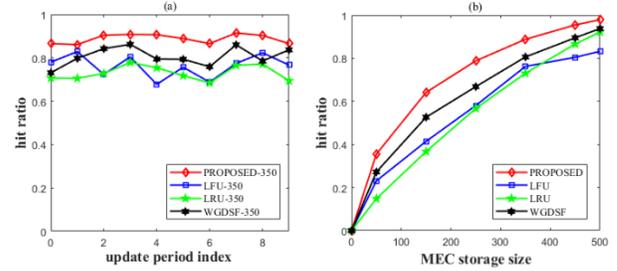

Fig.2. Hit ratios of different algorithm with the size of MEC server, $S=350$, $S=0\sim500$

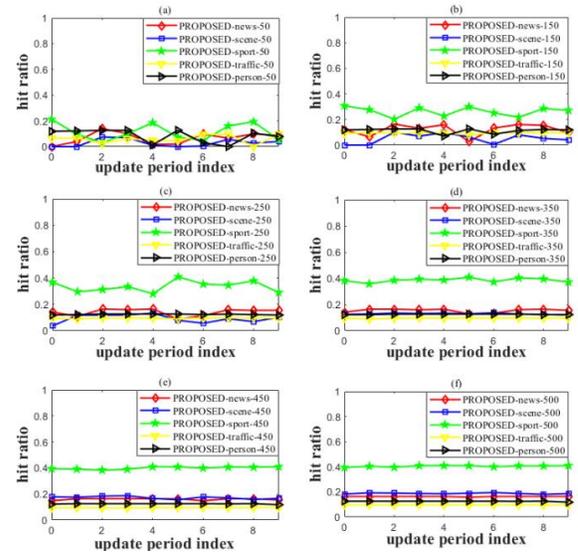

Fig.3. Hit ratios of different content types of PROPOSED with the size of MEC server $S=50,150,250,350,450,500$

Figure 3 shows the hit ratio of different video contents in different storage sizes of the MEC server. As is said in the introduction to Section III, PROPOSED will perform differential caching on the basis of the user's preference. According to the statistical analysis of all the user preferences, there are more users who prefer sports videos in the MEC coverage area. Therefore, PROPOSED can cache more sports videos to meet the needs of most users. As shown in this figure, with the increase of the storage capacity of MEC server, the hit rate of sports videos has increased significantly, and eventually become stable. The hit rate of sports videos has not increased significantly when the storage size of MEC server is 50 storage units. The reason is that the storage capacity of the MEC server is too small to represent the distribution of all the user preferences. The hit rate of all the videos has gradually stabilized when the storage size of MEC server is 350 storage units. This result occurs because of the fact that the storage capacity of MEC server is large enough to cache most videos, which is a good representation of all user preference distributions.

As shown in figure 3, PROPOSED is more inclined to cache sports videos because there are more users who like sports videos. Therefore, we analyze the differential cache performance of PROPOSED. As shown in Figure 4, compared with the LRU, LFU and WGDSF, PROPOSED shows the best performance for sports videos. With the increase of the storage capacity of the MEC server, PROPOSED shows that the hit rate of sports videos increases significantly, and eventually become stable. The specific reason is explained in Figure 3. As the hit ratio is increased, the backhaul network load and the startup delay are reduced, as shown in Figure 5.

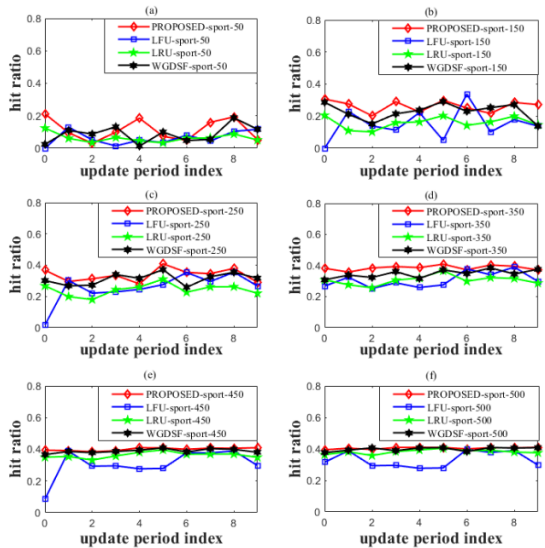

Fig.4. Hit ratios of all the sports videos of different algorithm with the size of MEC server $S$=50,150,250,350,450,500

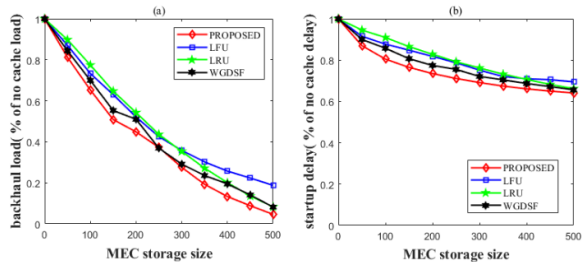

Fig.5. Backhaul load and startup delay

## 6. CONCLUSION

In this paper we consider the joint optimization problem of video caching and video real-time transcoding for video on demand service. Video popularity, user preferences, and video representations are taken into account in this optimization problem to guide the cache strategy of MEC server. This optimization problem is assumed to be a grouping knapsack problem. The dynamic programming is used to solve it in order to achieve an optimal solution. We provide numerical simulation results in terms of the cache hit ratio. These results show that our proposed algorithm can perform differential caching by analyzing user preferences, and achieves a better performance compared to the methods LRU, LFU and WGDSF.